\documentclass[preprint,superscriptaddress,preprintnumbers,showpacs]{revtex4}
\usepackage{amssymb}
\usepackage{graphicx}
\usepackage{dcolumn}
\usepackage{bm}

\begin{document}

\title{ New Parametrization of Neutrino Mixing Matrix}

\author{H. B. Benaoum}\email{hbenaoum@jlab.org, hbenaoum@pmu.edu.sa}\affiliation{Prince Mohammad Bin Fahd University, Al-Khobar 31952, Saudi Arabia } 

\date{October 20, 2010}

\begin{abstract} 
Global fits to neutrino oscillation data are compatible with tri-bimaximal mixing pattern, which predict $\theta_{23} = \frac{\pi}{4}, \theta_{12} = \sin^{-1} \left( \frac{1}{\sqrt{3}} \right)$ and $\theta_{13} = 0$. 
We propose here to parametrize the tri-bimaximal mixing matrix $V_{TBM}$ by its hermitian generator $H_{TBM}$ using the exponential map. Then we use the exponential map to express the deviations from tri-bimaximal pattern by deriving the hermitian matrices $H_{z=0}$ and $H_1$. These deviations might come from the symmetry breaking of the neutrino and charged lepton sectors. 
\end{abstract}


\maketitle

In recent years, masses and mixing of different generations of neutrinos have been the subject of numerous experimental and theoretical studies. These efforts not only confirmed the oscillations of neutrinos but also measured the mass-squared differences of the neutrino mass eigenstates.  \\
Neutrino oscillation was suggested by Pontecorvo \cite%
{pontecorvo} fifty years ago by analogy with the Kaon system  and further generalized to three flavors by Maki, Nakagawa and Sakata ( PMNS ) ~\cite{maki}. Visually, these mixings could be regarded as a rotation from fermion mass eigenstates to flavor eigenstates. The phenomenon of neutrino mixing can be simply described by the PMNS unitary matrix $V$,which links the neutrino flavor eigenstates $\nu_e, \nu_{\mu}, \nu_{\tau}$ to the mass eigenstates $\nu_1, \nu_2, \nu_2$ : \\

\begin{eqnarray} 
V  & = & \left(  \begin{array}{ccc} 
V_{e 1} & V_{e 2} & V_{e 3} \\
V_{\mu 1} & V_{\mu 2} & V_{\mu 3} \\
V_{\tau 1} & V_{\tau 2} & V_{\tau 3} \\ 
\end{array}
\right)
\end{eqnarray}
In the standard parametrization used by the Particle Data Group ( PDG ), the PMSN matrix is expressed by three mixing angles $\theta_{12}, \theta_{23}$ and $\theta_{13}$ and one intrinsic CP violating phase $\delta$ for Dirac neutrinos,  
\begin{eqnarray} 
V \left( \theta_{12}, \theta_{23}, \theta_{13}, \delta \right)  & = &  {  \left( 
\begin{array}{ccc}
c_{12} c_{13} & s_{12} c_{13} & s_{13} e^{- i \delta} \\
- s_{12} c_{23} - c_{12} s_{23} s_{13} e^{i \delta} & c_{12} c_{23} - s_{12} s_{23} s_{13} e^{i \delta} & s_{23} c_{13} \\
s_{12} s_{23} - c_{12} c_{23} s_{13} e^{i \delta} & - c_{12} s_{23} - s_{12} c_{23} s_{13} e^{i \delta} & c_{23} c_{13} \\  
\end{array} 
\right) } 
\end{eqnarray}
where $c_{ij} = \cos \theta_{ij}, s_{ij} = \sin \theta_{ij}$ and for Majorana neutrino, there are additional two independent Majorana CP violating phases given by a diagonal matrix 
$K = diag \left(1 , e^{\sigma_1}, e^{\sigma_2} \right)$. We will ignore the Majorana CP violating 
phase in this work. \\

The merit of this parametrization is that the three mixing angles are directly related to the mixing angles of solar, atmospheric and CHOOZ reactor neutrino oscillations, namely : 
\begin{eqnarray}
\theta_{12} & \approx & \theta_{sun} \nonumber \\
\theta_{23} & \approx & \theta_{atm} \nonumber \\
\theta_{13} & \approx  & \theta_{chz} 
\end{eqnarray}  

Moreover, the measure of CP violation is expressed through the rephasing invariant Jarlskog parameter defined as ~\cite{jarlskog} : 
\begin{eqnarray}
J & = & Im \left( V_{e 2} V_{\mu 3} V^{\star}_{e 3} V^{\star}_{\mu 2} \right)
\end{eqnarray} 

The exact expressions between the sines of the three mixing angles and the moduli of the elements of the PMNS mixing matrix are given by : 
\begin{eqnarray}
\sin^2 \theta_{13} & = & | V_{e 3} |^2 \nonumber \\
\sin^2 \theta_{12} & = & \frac{| V_{e 2} |^2}{1 - | V_{e 3} |^2} \nonumber \\
\sin^2 \theta_{23} & = & \frac{| V_{\mu 3} |^2}{1 - | V_{e 3} |^2} 
\end{eqnarray} 

Our current knowledge on the magnitude of the elements of the neutrino mixing matrix comes from the experiments on neutrino oscillations. A recent global $3 \nu$ oscillation analysis of neutrino experimental data gives the following numerical ranges ~\cite{fogli} :  \\
%\begin{center}
\begin{table}[!h]  
\begin{center}
\begin{tabular}{|c|c|c|c|} 
\hline
\hline  
Parameter & $\sin^2 \theta_{12}$ & $\sin^2 \theta_{23}$ & $\sin^2 \theta_{13}$ \\ 
\hline
Best fit & $0.312$ & $0.466$ & $0.016$ \\ 
\hline
$1 \sigma$ range & $0.294 -- 0.331$ & $0.408 –- 0.539$ & $0.006 –- 0.026$ \\ 
\hline
$2 \sigma$ range & $0.278 –- 0.352$ & $0.366 -– 0.602$ & $< 0.036$ \\ 
\hline
$3 \sigma$ range & $0.263 –- 0.375$ & $0.331 –- 0.644$ & $< 0.046$ \\
\hline 
\hline 
\end{tabular}
\caption{Global $3 \nu$ oscillation analysis with best fit values and allowed $n_{\sigma}$ ranges.}
\label{table1}  
\end{center}
\end{table}  
~\\

These experimental data are consistent with the so-called tri-bimaximal mixing (TBM) matrix pattern where the low energy neutrino mixing matrix $V_{TBM}$ ~\cite{harrison} is given by : 
\begin{eqnarray} 
V_{TBM} & = & R_{23} \left( x_0 \right) R_{12} \left( y_0 \right) \nonumber \\
& = &  ~\left( \begin{array}{ccc} 
\cos y_0 & \sin  y_0 & 0 \\
- \cos x_0 \sin y_0 & \cos x_0 \cos y_0 & \sin x_0 \\
\sin x_0 \sin y_0 & - \sin x_0 \cos y_0 & \cos x_0 \\
\end{array} \right)
\end{eqnarray} 
with $R_{23}$ and $R_{12}$ are rotations with angle $x_0 = \pi/4$ and 
$y_0 = \sin^{-1} \left( 1/\sqrt{3} \right)$.  \\
The TBM mixing has been studied extensively in the literature \cite{xing}. Parametrization of neutrino mixing matrix with TBM as the zeroth order has been considered by several authors with different approaches \cite{pakvasa}. Deviations from $\theta_{13} = 0$ due to charged lepton corrections \cite{king} and renormalization group running effects \cite{antusch} have been considered in literature. \\

In this paper, we propose to parametrize  the TBM mixing matrix $V_{TBM}$ by its generator $H_{TBM}$ using the exponential map. Indeed, in Lie theory of groups and their corresponding algebras the exponential map is a crucial tool because it gives the connection between Lie algebra element and their corresponding Lie group element. Using this connection, we derive the generator $H_{z=0}$ of the mixing matrix $V_{z=0}$. Then we use the exponential map to find the hermitian generator $H_1$ which expresses the deviations from $z = \theta_{13} =0$. 
For us, both hermitian matrices $H_{z=0}$ and $H_1$ encode deviations from TBM hermitian generator $H_{TBM}$ since there is no convincing reason for TBM to be exact. Moreover, we consider that the 
$H_{z=0}$ arises from the diagonalization of the neutrino mass matrix $M_{\nu}$ whereas $H_1$ comes from the diagonalization of the charged lepton sector 
$M_l$.  \\ 
 
The unitary TBM matrix has three distincts eigenvalues $\lambda^0_i$ as follows :  
\begin{eqnarray}
\lambda^0_1 & = & 1 \nonumber \\
\lambda^0_{2,3} & = & \frac{Tr ( V_{TBM} ) -1 \pm i N_0}{2}
\end{eqnarray}
Here $N_0$ is a normalizing factor which can be expressed in terms of the trace of TBM mixing matrix as : 
\begin{eqnarray}
N_0 & = & \sqrt{4 - \left( Tr ( V_{TBM} ) -1 \right)^2}
\end{eqnarray}

Based on this we write $V_{TBM}$ in terms of its eigenvalues as : 
\begin{eqnarray}
V_{TBM} & = & U^{\dagger} ~\Lambda_0~ U 
\end{eqnarray}
where $\Lambda_0 = diag \left( \lambda^0_1, \lambda^0_2, \lambda^0_3 \right)$ and the diagonalizing unitary 
matrix $U$ is given by : 
\begin{eqnarray}
U & = & \frac{1}{N_0}~ \left( \begin{array}{ccc}
\sin x_0  \left( 1+ \cos y_0 \right) &  
\sin x_0 \sin y_0 & 
\sin y_0 \left(1 + \cos x_0 \right) \\
- \sqrt{2} ~\left( \lambda^0_2 + 1 \right) \cos ( \frac{y_0}{2} ) & 
 \sqrt{2} ~\left( \lambda^0_2 - \cos x_0 \right) \cos ( \frac{y_0}{2} ) & 
\sqrt{2} ~\sin x_0 \cos ( \frac{y_0}{2} ) \\  
 - \sqrt{2} ~\left( \lambda^0_3 + 1 \right) \cos ( \frac{y_0}{2} ) & 
 \sqrt{2} ~\left( \lambda^0_3 - \cos x_0 \right) \cos ( \frac{y_0}{2} ) & 
\sqrt{2} ~\sin x_0 \cos ( \frac{y_0}{2} ) \\  
\end{array} \right)
\end{eqnarray}

Now the diagonal matrix $\Lambda_0 = diag \left( \lambda^0_1, \lambda^0_2, \lambda^0_3 \right)$ which has three roots of unity can be rewritten as :
\begin{eqnarray}
\Lambda_0 & = e^{i~\Omega_0}
\end{eqnarray}
where $\Omega$ is given by : 
\begin{eqnarray}
\Omega_0 & = & \left( \begin{array}{ccc} 
0 & 0 & 0 \\
0 &  \omega_0 & 0 \\
0 & 0 & -\omega_0 \\
\end{array} \right) 
\end{eqnarray}
and $\omega_0 = -i ~\ln \left( \lambda^0_2 \right) = \arctan \left( \frac{N_0}{Tr ( V_{TBM}) -1} \right)$ .  \\

With this approach, we can express the TBM mixing matrix $V_{TBM}$ using the exponential map as : 
\begin{eqnarray}
V_{TBM} & = & U^{\dagger}~ e^{i~\Omega_0} ~U  \nonumber \\
& = & e^{i H_{TBM}} 
\end{eqnarray}
where the Hermitian matrix $H_{TBM}$ is : 
\begin{eqnarray}
H_{TBM} & = & U^{\dagger} ~\Omega~ U  \nonumber \\
& = & \frac{i \omega_0}{N_0} ~\left( \begin{array}{ccc}
0 & - \left(1 + \cos x_0 \right) \sin y_0 & \sin x_0 \sin y_0 \\
 \left(1 + \cos x_0 \right) \sin y_0 & 0 & - \sin x_0 \left( 1 + \cos y_0 \right) \\
- \sin x_0 \sin y_0 & \sin x_0 \left( 1 + \cos y_0 \right) & 0 \\
\end{array} \right)
\end{eqnarray} 
As we know, the exponential map connects Lie algebra element to their corresponding Lie group element. Here $H_{TBM}$ is the generator for the TBM mixing matrix. \\

In this parametrization, the Hermitian matrix $H_{TBM}$ gives the zeroth order of the neutrino mixing matrix 
and for TBM values $x_0 = \frac{\pi}{4}$ and $y_0 = \sin^{-1} \left( \frac{1}{\sqrt{3}} \right)$, it has the 
following simple expression :
\begin{eqnarray}
H_{TBM} & = & i 
\frac{\arctan \left(
\frac{{\sqrt{4 - \left(-1 \sqrt{\frac{2}{3}} +\frac{1}{\sqrt{2}} + \frac{1}{\sqrt{3}} \right)^2}}}
{-1 + \sqrt{\frac{2}{3}} + \frac{1}{\sqrt{2}} + \frac{1}{\sqrt{3}}} \right) }
{\sqrt{4 - \left(-1 \sqrt{\frac{2}{3}} +\frac{1}{\sqrt{2}} + \frac{1}{\sqrt{3}} \right)^2}} 
\left( \begin{array}{ccc} 
0 & - ( \frac{1}{\sqrt{3}} + \frac{1}{\sqrt{6}} ) & \frac{1}{\sqrt{6}} \\
\frac{1}{\sqrt{3}} + \frac{1}{\sqrt{6}} & 0 & - ( \frac{1}{\sqrt{2}} + \frac{1}{\sqrt{3}} ) \\
- \frac{1}{\sqrt{6}} & \frac{1}{\sqrt{2}} + \frac{1}{\sqrt{3}} & 0 \\
\end{array}  \right) \nonumber \\
& = &  i~\left( \begin{array}{ccc} 
0 & - 0.5831 & 0.2415 \\
0.5831 & 0 & - 0.7599 \\
- 0.2415 & 0.7599 & 0 \\
\end{array} \right)
\end{eqnarray}

The above parametrization of TBM ( 14 ) is written in terms of mixing angles $x_0$ and $y_0$, but it is easy to see that the elements of the Hermitian matrix $H_{TBM}$ are related directly to the elements of the mixing matrix $V_{TBM}$ as : 
\begin{eqnarray}
H_{TBM} & = & \frac{i \omega_0}{N_0} ~ \left( \begin{array}{ccc} 
0 & V_{TBM~21} -V_{TBM~12} & V_{TBM~31} -V_{TBM~13} \\
- V_{TBM~21} + V_{TBM~12}  & 0 & V_{TBM~32} -V_{TBM~23} \\
- V_{TBM~31} + V_{TBM~13}  & - V_{TBM~32} + V_{TBM~23}  & 0 \\
\end{array} \right) 
\end{eqnarray}
where $\omega_0$ and $N_0$ are expressed by the invariant parameter $Tr \left( V_{TBM} \right) = V_{TBM~11} + V_{TBM~22} + V_{TBM~33}$. \\

Many future experiments are geared towards improving the precision measurements of the mixing angles and mass-squared differences, so departure from the exact tri-bimaximal mixing is expected. 
The signature of the deviation from the TBM mixing scenario comes from the measurement of a non-zero value for $V_{e 3}$ and/or the observation of CP violation in neutrino oscillations. \\

The deviation from TBM mixing scheme can be parametrized by writing the PMNS matrix $V$ in terms of the 
exponential map $V_{z=0} = e^{i H_{z=0}}$ as : 
\begin{eqnarray}
V \left(x,y,z, \delta \right) & = & V_{z =0} ~V_1
\end{eqnarray}
where the Hermitian matrix $H_{z=0}$ has the same form as $H_{TBM}$,
\begin{eqnarray}
H_{z=0} & = &  \frac{i \omega}{N} ~\left( \begin{array}{ccc}
0 & - \left(1 + \cos x \right) \sin y & \sin x \sin y \\
 \left(1 + \cos x \right) \sin y & 0 & - \sin x \left( 1 + \cos y \right) \\
- \sin x \sin y & \sin x \left( 1 + \cos y \right) & 0 \\
\end{array} \right)
\end{eqnarray} 
with $\omega_{z=0} = \arctan \left( \frac{N_{z=0}}{Tr ( V_{z=0}) -1} \right)$ and $N_{z=0} = \sqrt{4 - \left( Tr ( V_{z=0} ) -1 \right)^2}$ .  \\
Similarly, we can express the elements of $H_{z=0}$ in terms of the observables of the mixing matrix $V_{z=0}$. We emphasize here that the parametrization in terms of the generator is completely general and is not based on the TBM pattern or $z=0$ ( {\em see Appendix where a more general result of the generator of the mixing matrix with no CP is found and written in terms of observables }). \\

Both $V_{z= 0}$ and $V_1$ encode deviations from tri-bimaximal mixing. Using the exponential map, the matrix $V_1 = e^{i H_1}$ generated by the Hermitian matrix $H_1$ is given by : 
\begin{eqnarray}
H_1 & = &  i~\left( \begin{array}{ccc} 
0 & 0 & - e^{-i \delta} z \cos y  \\
0 & 0 & - e^{-i \delta} z \sin y  \\
e^{i \delta} z \cos y  & e^{i \delta} z \sin y  & 0 \\
\end{array} \right) 
\end{eqnarray}
CP violation is described by the Jarlskog invariant which to third order in z is : 
\begin{eqnarray}
J & = & \frac{1}{4} z \left(1 - \frac{7}{6} z^2 \right) \sin 2 x \sin 2 y \sin \delta + O \left( z^4 \right)
\end{eqnarray} 

The best fit values for $x= \theta_{23}, y = \theta_{12}$ and $z = \theta_{13}$ of the global analysis (see table above) leads to : 
\begin{eqnarray}
H_{z=0} & = & i~\left( \begin{array}{ccc} 
0 & - 0.5642 & 0.2225 \\
0.5642 & 0 & - 0.7288 \\
- 0.2225 & 0.7288 & 0 \\
\end{array} \right)
\end{eqnarray} 
and 
\begin{eqnarray}
H_1 & = & i~\left( \begin{array}{ccc} 
0 & 0& -0.1052 ~e^{- i \delta} \\
0 & 0 & - 0.0160 ~e^{- i \delta} \\
0.1052 ~e^{ i \delta} & 0.0160 ~e^{i \delta} & 0 \\
\end{array} \right)
\end{eqnarray} 
The Jarlskog invariant is : 
\begin{eqnarray}
J = 0.02876 \sin \delta
\end{eqnarray}

We may consider that $V_{z=0}$ which has TBM structure is coming from neutrino sector and $V_1$ coming from 
symmetry breaking in the charged lepton sector. The neutrino mass matrix is given as :  
\begin{eqnarray}
M_{\nu}  & =  &
V_{z=0} ~diag \left( m_{\nu_1}, m_{\nu_2}, m_{\nu_3} \right)~ V_{z=0}^{T}  \\ 
& = & {\small \left( \begin{array}{ccc}
m_{\nu_1} + 2 \Delta_{\nu_{21}} \sin^2 y  & \Delta_{\nu_{21}} \cos x \sin 2y & - \Delta_{\nu_{21}} \sin x \sin 2 y \\
\Delta_{\nu_{21}} \cos x \sin 2y & m_{\nu_2} - 2 \Delta_{\nu_{21}} \cos^2 x \sin^2 y + 2 \Delta_{\nu_{32}} \sin^2 x  & 
\Delta_{\nu_{21}} \sin 2 x \sin^2 y + m_{\nu_{23}} \sin 2 x \\
- \Delta_{\nu_{21}} \sin x \sin 2 y & \Delta_{\nu_{21}} \sin 2 x \sin^2 y + m_{\nu_{23}} \sin 2 x & 
m_{\nu_3} - 2 \Delta_{\nu_{21}} \sin^2 x \sin^2 y - 2 \Delta_{\nu_{32}} \sin^2 x \\
\end{array} \right) } \nonumber 
\end{eqnarray} 	
Here we have defined $m_{\nu_{ij}} = \frac{m_{\nu_i} + m_{\nu_i}}{2}$ and $\Delta_{\nu_{ij}} = \frac{m_{\nu_i} - m_{\nu_i}}{2}$ . \\

To first order in $z$, the charged lepton mass matrix is : 
\begin{eqnarray} 
M_l & = & V_1 ~diag \left( m_e, m_{\mu}, m_{\tau} \right)~ V_1^{\dagger}  \\ 
& = & \left( \begin{array}{ccc} 
m_e & 0 & e^{- i \delta} \Delta_{\tau e}  ~z \cos y \\
0 & m_{\mu} & e^{- i \delta} \Delta_{\tau \mu} ~z \sin y \\
e^{i \delta} \Delta_{\tau e} ~z \cos y & e^{i \delta} \Delta_{\tau \mu} ~z \sin y & m_{\tau} \\
\end{array} \right) + O ( z^2 ) \nonumber 
\end{eqnarray}
where we defined $\Delta_{ij} = \frac{m_i - m_j}{2}$ with $i = e,\mu, \tau$ . \\
\begin{acknowledgments}
I would like to thank S. Nasri, W. Rodejohann and J. Schechter for reading the manuscript and their useful comments. 
\end{acknowledgments}

\appendix 
\begin{center}
{\bf Appendix}
\end{center}
We present here a general result for the connection of the Lie group element $V$ and Lie algebra element $H$ without CP phase. \\
The mixing matrix whitout CP is given by : 
\begin{eqnarray}
V_{\delta = 0} & = & \left( \begin{array}{ccc} 
\cos y \cos z & \sin y \cos z & \sin z \\
- \cos x \sin y - \sin x \cos y \sin z & \cos x \cos y - \sin x \sin y \sin z & \sin x \cos z \\
\sin x \sin y - \cos x \cos y \sin z & - \sin x \cos y - \cos x \sin y \sin z & \cos x \cos z \\
\end{array} \right) 
\end{eqnarray}
The unitary matrix $V_{\delta=0}$ has three distincts eigenvalues $\lambda_i$ as follows :  
\begin{eqnarray}
\lambda_1 & = & 1 \nonumber \\
\lambda_{2,3} & = & \frac{Tr ( V_{\delta=0} ) -1 \pm i N_{\delta=0}}{2}
\end{eqnarray}
Here $N_{\delta=0}$ is a normalizing factor which can be expressed in terms of the trace of mixing matrix $V_{\delta=0}$ as : 
\begin{eqnarray}
N_{\delta=0} & = & \sqrt{4 - \left( Tr ( V_{\delta=0} ) -1 \right)^2}
\end{eqnarray}
Using the above approach, it is easy to find the matrix $V_{\delta=0}$ in terms of the exponential map as : 
\begin{eqnarray}
V_{\delta=0} & = & e^{i ~H_{\delta =0}} 
\end{eqnarray}
where the Hermitian generator matrix can be expressed in terms of the elements of the mixing matrix $V_{\delta=0}$ as : 
\begin{eqnarray}
H_{\delta =0} & = & \frac{i \omega_{\delta=0}}{N_{\delta=0}} ~ \left( \begin{array}{ccc} 
0 & V_{\delta=0,21} -V_{\delta=0,12} & V_{\delta=0,31} -V_{\delta=0,13} \\
- V_{\delta=0,21} + V_{\delta=0,12}  & 0 & V_{\delta=0,32} -V_{\delta=0,23} \\
- V_{\delta=0,31} + V_{\delta=0,13}  & - V_{\delta=0,32} + V_{\delta=0,23}  & 0 \\
\end{array} \right) 
\end{eqnarray}
where $\omega_{\delta=0}$ and $N_{\delta=0}$ are expressed by the invariant parameter, 
\begin{eqnarray}
Tr \left( V_{\delta=0} \right) = V_{\delta=0,11} + V_{\delta=0,22} + V_{\delta=0,33}
\end{eqnarray}
$V_{\delta=0}$ is given by : 
\begin{eqnarray}
V_{\delta=0} & = & {\bf 1_{3x 3}} + i \frac{\sin \omega_{\delta=0}}{\omega_{\delta=0}} ~H_{\delta=0} - 
\frac{1 - \cos \omega_{\delta=0}}{\omega_{\delta=0}^2} ~H^2_{\delta=0} 
\end{eqnarray}

By using best fit, we get :
\begin{eqnarray}
H_{\delta=0} & = & i ~\left( \begin{array}{ccc}
0 & -0.611 & 0.105 \\
0.611 & 0 & - 0.765 \\
- 0.105 & 0.765 & 0 \\
\end{array} \right) 
\end{eqnarray}

\end{document}